\documentclass[twocolumns]{aa}
\usepackage[varg]{txfonts}
\usepackage{cancel}
\usepackage{natbib}
\usepackage{textcomp}
\bibpunct{(}{)}{;}{a}{}{,} 

\usepackage[backref]{hyperref}
\usepackage{url}
\usepackage[displaymath,mathlines]{lineno}

\usepackage{xcolor}
\usepackage{graphicx}
\usepackage{subfigure}
\definecolor{paolo}{rgb}{0.0, 0., 1.0}

\title{Connecting Solar Orbiter remote-sensing observations and Parker Solar Probe in-situ measurements with a numerical MHD reconstruction of the Parker spiral}
\titlerunning{Connecting Solar Orbiter and PSP with a numerical MHD reconstruction of the Parker spiral}
\authorrunning{R. Biondo et al.}
\author{Ruggero {Biondo}\inst{\ref{UNIPA},\ref{OATO}}
\and Alessandro Bemporad\inst{\ref{OATO}}
\and Paolo Pagano\inst{\ref{UNIPA},\ref{OAPA}}
\and Daniele Telloni\inst{\ref{OATO}}
\and Fabio Reale\inst{\ref{UNIPA},\ref{OAPA}}
\and Marco Romoli\inst{\ref{UNIFI}}
\and Vincenzo Andretta\inst{\ref{OACN}}
\and Ester Antonucci\inst{\ref{OATO}}
\and Vania Da Deppo\inst{\ref{CNR-IFN}}
\and Yara De Leo\inst{\ref{UNICT},\ref{MaxPlanck}}
\and Silvano Fineschi\inst{\ref{OATO}}
\and Petr Heinzel\inst{\ref{Czech}}
\and Daniel Moses\inst{\ref{NASA}}
\and Giampiero Naletto\inst{\ref{UNIPD},\ref{CNR-IFN}}
\and Gianalfredo Nicolini\inst{\ref{OATO}}
\and Daniele Spadaro\inst{\ref{OACT}}
\and Marco Stangalini\inst{\ref{ASI}}
\and Luca Teriaca\inst{\ref{MaxPlanck}}
\and Federico Landini\inst{\ref{OATO}}
\and Clementina Sasso\inst{\ref{OACN}}
\and Roberto Susino\inst{\ref{OATO}}
\and Giovanna Jerse\inst{\ref{OATS}}
\and Michela Uslenghi\inst{\ref{IASF}}
\and Maurizio Pancrazzi\inst{\ref{OATO}}
}

\institute{University of Palermo, Physics and Chemistry Department, Piazza del Parlamento 1, I-90134 Palermo, Italy\label{UNIPA}
\and INAF-Turin Astrophysical Observatory, via Osservatorio 20, I-10025 Pino Torinese (TO), Italy\label{OATO}
\and INAF-Palermo Astronomical Observatory, Piazza del Parlamento 1, I-90134 Palermo, Italy\label{OAPA}
\and University of Firenze, Department of Physics and Astronomy, Via Giovanni Sansone 1, I-50019 Sesto Fiorentino, Italy\label{UNIFI}
\and INAF-Capodimonte Observatory, Salita Moiariello 16, I-80131 Napoli, Italy\label{OACN}
\and CNR-Institute for Photonics and Nanotechnologies, Via Trasea 7, I-35131 Padova, Italy, Italy\label{CNR-IFN}
\and Czech Academy of Sciences, Astronomical Institute, Fričova 298, CZ-25165 Ondrejov, Czech Republic\label{Czech}
\and NASA, Headquarters, Washington, DC 20546, USA\label{NASA}
\and University of Padova, , Department of Physics and Astronomy, Via Francesco Marzolo 8, I-35131 Padova, Italy\label{UNIPD}
\and INAF-Catania Astrophysical Observatory, Via S. Sofia 78, I-95123 Catania, Italy\label{OACT}
\and University of Catania, Department of Physics and Astronomy, Via Santa Sofia 64, I-95123 Catania, Italy\label{UNICT}
\and Italian Space Agency, Via del Politecnico snc, I-00133 Roma, Italy\label{ASI}
\and Max Planck Institute for Solar System Research, Justus-von-Liebig-Weg 3, D-37077 Göttingen, Germany\label{MaxPlanck}
\and INAF-Trieste Astronomical Observatory, Località Basovizza 302, I-34149 Trieste, Italy\label{OATS}
\and INAF-Institute of Space Astrophysics and Cosmic Physics of Milan, Via Alfonso Corti 12, I-20133 Milano, Italy\label{IASF}
}

\begin{document}

\abstract{As a key feature, NASA’s Parker Solar Probe (PSP) and ESA-NASA’s Solar Orbiter (SO) missions cooperate to trace solar wind and transients from their sources on the Sun to the inner interplanetary space.
The goal of this work is to accurately reconstruct the interplanetary Parker spiral and the connection between coronal features observed remotely by the Metis coronagraph on-board SO and those detected in situ by PSP at the time of the first PSP-SO quadrature of January 2021.
We use the Reverse In-situ and MHD Approach (RIMAP), a hybrid analytical-numerical method performing data-driven reconstructions of the Parker spiral. RIMAP solves the MHD equations on the equatorial plane with the PLUTO code, using the measurements collected by PSP between 0.1 and 0.2 AU as boundary conditions. Our reconstruction connects density and wind speed measurements provided by Metis (3-6 solar radii) to those acquired by PSP (21.5 solar radii) along a single streamline.
The capability of our MHD model to connect the inner corona observed by Metis and the super Alfvénic wind measured by PSP, not only confirms the research pathways provided by multi-spacecraft observations, but also the validity and accuracy of RIMAP reconstructions as a possible test bench to verify models of transient phenomena propagating across the heliosphere, such as coronal mass ejections, solar energetic particles and solar wind switchbacks.}

\maketitle

\section{Introduction}\label{sec:Intro}
The number of spacecraft dedicated to the study of the solar atmosphere, the heliosphere and space weather have significantly increased in the last few years. They shed light on open questions related to the emergence of the solar magnetic field, the generation of the solar wind, and the acceleration of energy particles. NASA's Parker Solar Probe \citep[PSP;][]{fox2016} and ESA-NASA's Solar Orbiter \citep[][]{muller2020} play a pivotal role in this. The former is the very first spacecraft to probe the solar corona with in situ measurements. The latter is the closest-to-the-Sun spacecraft to provide both in situ and remote sensing measurements of the Sun, with the aim to accurately describe the magnetic connectivity between the solar wind plasma and its source regions at the Sun. A key feature of both missions is their cooperation, which allows the tracking of solar wind and related transients from their sources on the Sun to the inner interplanetary space \citep{velli2020AA}.

Observations of the same solar wind plasma parcel at two different heliocentric distances were first attempted from an alignment between the two Helios probes \citep{schwartz_marsch_1983}. Successive alignments in the inner and outer heliosphere were exploited to study the evolution of solar wind turbulence \citep[e.g.][]{damicis2010,bruno2014} and interplanetary coronal mass ejections \citep{telloni2020_magneticcloud}. In particular, the first PSP-Solar Orbiter lineup was recently investigated by \cite{telloni2021_radial}.

Additionally, combinations of remote sensing observations of the solar wind near the Sun and in situ measurements far from the Sun were first attempted during the so-called SOHO-Ulysses quadratures \citep{Suess2000, Suess2001, Poletto2002}. The UltraViolet Coronagraph Spectrometer (UVCS) allowed us to compare not only the different wind stream velocities, but also the elemental abundances and the so-called First Ionization Potential (FIP) effect \citep{Bemporad2003}. From these quadratures we could also study for the first time the same plasma observed during and after solar eruptions with remote sensing and in situ data \citep{Suess2004, Poletto2004, Bemporad2006, Owens2008}. Later on, with the Solar TErrestrial RElations Observatory (STEREO) mission, combinations of remote sensing and in situ data analysis focused on transient events \citep{Mostl2009,  Kilpua2009, Innes2010}. Attempts to identify the solar wind sources on the Sun based on abundance measurements were also carried out with the Extreme-ultraviolet Imaging Spectrometer (EIS) on board Hinode \citep{Brooks2011, Brooks2015}.

All these previous works made various assumptions to reconstruct the path followed by the solar plasma propagating from the inner corona to the interplanetary medium. The goal of the present work is to accurately reconstruct the Parker spiral and to connect the coronal features observed remotely by the Metis coronagraph \citep{antonucci2020} on board Solar Orbiter with those detected in situ by PSP, at the time of the first PSP-Solar Orbiter quadrature which occurred on January 2021 \citep{telloni2021}, by using the Reverse In-situ and Magnetohydrodynamics APproach \citep[RIMAP, ][]{Biondoetal2021}, a hybrid analytical-numerical method for reconstructing  the Parker spiral. 

\section{Data and model}\label{sec:Model}
The PSP measurements are used here as boundary conditions for the RIMAP backward reconstruction. Specifically, the PSP data refer to the time interval January 15-21, 2021, during solar encounter \#$7$.
The magnetic field and plasma measurements come from the fluxgate magnetometer of the Electromagnetic Fields Investigation (FIELDS) suite \citep{bale2016} and the Solar Probe Analyzers A for ions (SPAN-Ai), top-hat electrostatic analyser of the Solar Wind Electrons Alphas and Protons suite \citep[SWEAP;][]{kasper2016}, respectively, and averaged at $1$ minute resolution.

\begin{figure}[htbp]
\centering
    \centering
    \includegraphics[width=\linewidth,trim={0cm 1.15cm 0cm 1.95cm},clip]{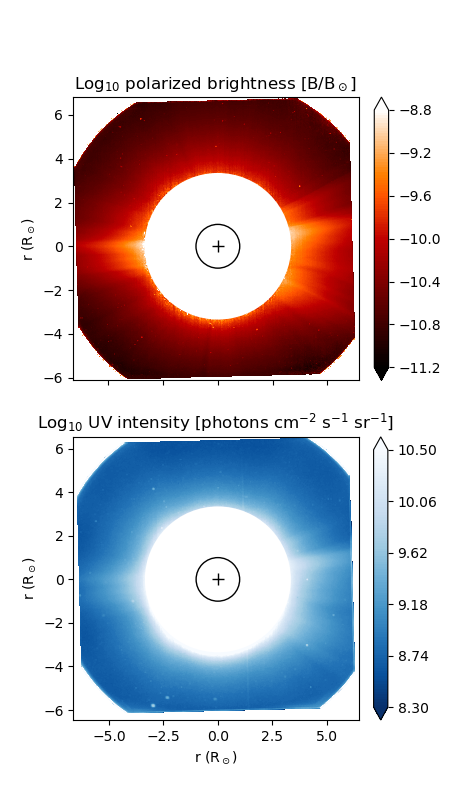}
    \caption{Metis coronal images of VL polarized brightness (top) and UV HI Ly-alpha (bottom) in the solar corona acquired on January 17, 2021, from 16:30 to 17:00 UT within the 3.5-6.3 R$_\odot$ field of view.}
    \label{fig:Metis_pBUV}
\end{figure}

Solar Orbiter reached perihelion on January 17, 2021 at 17:39:21 UT at a heliocentric distance equal to 0.583 AU. Close to perihelion, when the spacecraft was at 188.9\textdegree\ longitude from the Sun-Earth line and 91.2\textdegree\ colatitude (anti-clockwise from the north pole), Metis obtained simultaneous images of the corona in the visible light (VL) channel between 580 and 640 nm and in the UV channel around the resonantly scattered HI Ly-$\alpha$ line at 121.6 nm \citep{antonucci2020}. One polarized-brightness sequence of four polarimetric images and two UV images with detector integration times of 30 and 60 s, respectively, were acquired in the time interval from 16.30 to 17.00 UT. Figure \ref{fig:Metis_pBUV} shows the VL polarized brightness and UV HI Ly-alpha coronal images of the solar corona in an annular field of view between 3.5 and 6.3 solar radii. The reduction of the Metis data follows the steps given in \citet{romoli2021}. As anticipated in \citet{antonucci2020}, the final radiometric calibration was verified and updated by using calibration stars. The additional UV channel correction steps, such as the correction of the spatial response disuniformity, were performed according to \citet{andretta2021}.
RIMAP \citep{Biondoetal2021} is a technique for reconstructing the Parker spiral on the equatorial plane, starting from an analytically built inner boundary and propagating it outwards by solving the time-dependent magnetohydrodynamics (MHD) equations in a frame corotating with the solar equator. 
This approach was shown to reproduce the fine longitudinal structure of the spiral and to accurately match the features measured at 1 AU. 
The RIMAP output can be used as background for the propagation of space weather transients such as interplanetary coronal mass ejections \citep{biondoetal2021_ICME}.
Here, we used RIMAP to connect observed coronagraphic observation from Metis to the in situ data from PSP.
The model is constrained by in situ measurements of plasma density, speed, and magnetic field taken by PSP between 0.1 and 0.2 AU. RIMAP numerically solves the MHD equations with the PLUTO code \citep{mignone2007,mignone2012}, starting from the in situ parameters analytically back-mapped as inner boundary conditions. The simulation reconstructs the Parker spiral from 5 to 60 solar radii, thus connecting the Metis field of view (3-6 solar radii) to the PSP orbit.

In our RIMAP framework, the computational grid extends from 5 R$_\odot$ to 60 R$_\odot$ and from -30\textdegree\ to 172\textdegree\ in the longitudinal domain (in the heliocentric inertial coordinate frame), while the colatitude is restricted to a 2\textdegree\ band centred around 90\textdegree (around the equatorial plane). The grid has 256, 8, and 1024 equal cells in the radial, latitudinal and longitudinal direction, thus maintaining a regular cell aspect since the extension in longitude is more than $\pi$ times larger than in radius. Although the Metis field of view is not entirely within the computational domain, it is far enough from the solar surface to exclude complex and transient structures, which would affect the reconstruction of a steady Parker spiral. The domain largely includes the PSP orbit, which extends to about 30 R$_\odot$ at most in our longitudinal range (see Fig. \ref{fig:RIMAP_eclipticmaps}).
Using the Parker spiral equation, we connected each PSP data point to a location in the RIMAP boundary at 5 R$_\odot$ through a streamline. When streamlines cross one another, one is removed \citep{Biondoetal2021} such that each location at the inner boundary is connected to a unique PSP data point and its density is computed from the continuity law along the selected streamline. At the inner boundary, the only non-zero speed component is the radial component, which is the one on the corresponding PSP data point reduced by 12\% to compensate for the simulated acceleration along the streamlines outwards in the less dense medium. The $\varphi$ velocity is null in the steady regime. The plasma temperature is rescaled as $T_\mathrm{PSP}\left(R_\mathrm{PSP}/5 R_\odot\right)^{0.5}$, and the radial and longitudinal magnetic field components are rescaled according to the model by \cite{weberdavis1967}. The latitudinal component of the magnetic field is instead set equal to zero. These scaled plasma and magnetic field parameters are used as inner steady boundary conditions to solve the time-dependent MHD equations with the PLUTO code. These conditions progressively build outwards a coherent Parker spiral, as in the RIMAP concept, until a steady condition is reached throughout the computational domain. 

\section{Results}\label{sec:Results}

\begin{figure*}[htbp]
    \centering
    \includegraphics[width=\linewidth,trim={1.2cm 0.6cm 1.2cm 0.9cm}]{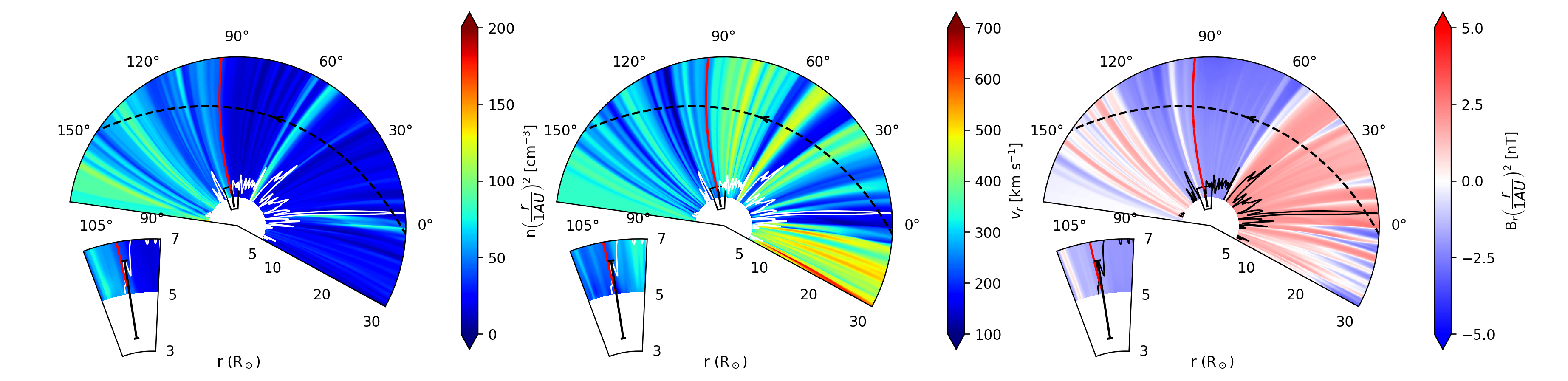}
    \caption{Equatorial maps of plasma density, radial speed and magnetic field as simulated by RIMAP, starting from the PSP data collected from January 15 to 21, 2021. The dashed black line is the PSP trajectory (projected on the equatorial plane), and the black segment from 3.5 to 6.3 R$_\odot$ is drawn at the Metis' plane-of-the-sky latitude on January 17. The solid red line is the field line connecting PSP measurements and Metis observations. The white line (black in the third panel) represents the Alfvén point for each streamline.}
    \label{fig:RIMAP_eclipticmaps}
\end{figure*}

Figure \ref{fig:RIMAP_eclipticmaps} shows the RIMAP-reconstructed Parker spiral (maps of density, velocity, and radial components of the magnetic field), where we mark the PSP equatorial trajectory from which the data were extracted (dashed black arrowed line), the longitude for the Metis plane-of-sky (in the inset, at 99\textdegree\ from 3.5 to 6.3 R$_\odot$), and the field line passing through the intersection between the PSP trajectory and Metis longitude (solid red line). A close-up of the Metis field of view is also shown in the lower left of each map. As discussed in \cite{Biondoetal2021}, the RIMAP technique allows us to reproduce the fine structures of the Parker spiral down to a size of 0.2 R$_\odot$, which is sufficient as each of the streamlines extracted from the PSP data is individually discernible. Most of the streamlines in the computational domain have low density ($n(r/\text{1 AU})^2\leq50\text{ cm}^{-3}$), with the $\varphi<90$\textdegree\ portion of the domain being emptier than the other one; exceptions include at -15\textdegree, the two close streamlines between 45\textdegree\ and 60\textdegree, and the ones around 90\textdegree\ that almost reach 150 cm$^{-3}$. The solar wind speed is overall anti-correlated with the density according to the continuity equation, with values close to the average speed of 400 km/s. The lowest speed ($\approx$100 km/s) is found in a streamline between 90\textdegree\ and 110\textdegree, while the highest (slightly less than 700 km/s) is found in the streamlines close to the initial longitudinal boundary of the model, between -30\textdegree\ and -15\textdegree.
The most striking feature of the radial component of the magnetic field is the inversion of direction at 50\textdegree, which is measured by PSP as it crosses the heliospheric current sheet (HCS).

Relying on the RIMAP-modelled plasma and magnetic field estimates, it is possible to assess for each streamline the Alfvén point, that is, the location where the solar wind speed equals that of the Alfvén waves. This is displayed as a function of the longitude in Fig. \ref{fig:RIMAP_eclipticmaps} and provides a very interesting piece of information, marking where the wind becomes super-Alfvénic (i.e. the outermost boundary of the solar corona).

\begin{figure*}[htbp]
\begin{center}
\includegraphics[width=\linewidth,trim={2.15cm 0.85cm 3cm 1.0cm},clip]{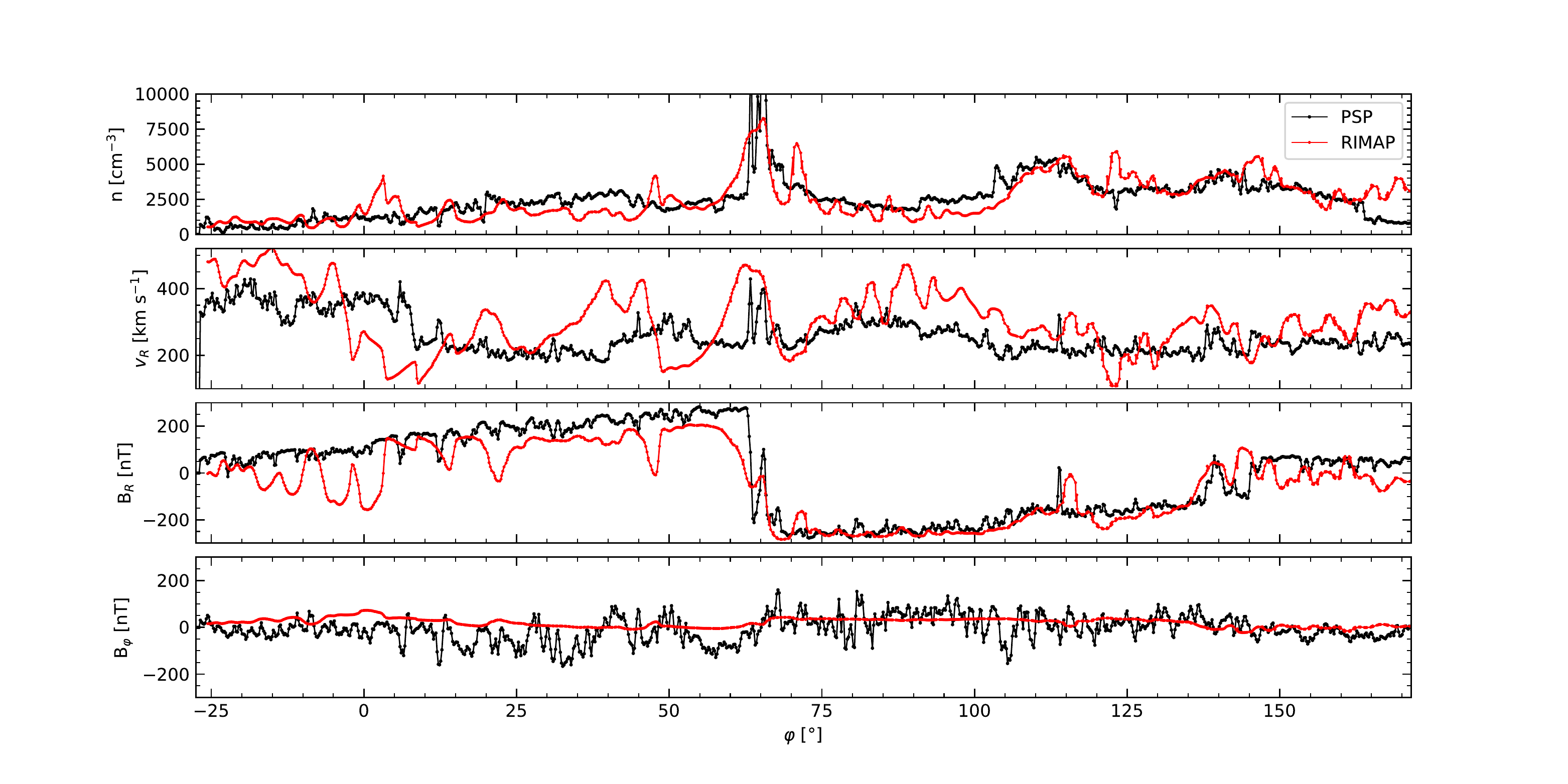}
\end{center}
\caption{Comparison between PSP-measured (black) and RIMAP-modelled (red) solar wind parameters from January 15-21, 2021. From top to bottom, the panels display the proton number density, the solar wind speed, the radial, and the tangential components of the interplanetary magnetic field (RTN).}
\label{fig:psp}
\end{figure*}

In Fig.\ref{fig:psp} we scan the computational domain reproducing the trajectory of PSP in this perihelion transit. We then compare relevant solar wind measurements by PSP on this scan (black) to the RIMAP results (red). We show, from top to bottom, the proton number density, the solar wind bulk speed, and the radial and longitudinal magnetic field components in the radial tangential normal (RTN) coordinate system taken along the PSP trajectory. 

Remarkably, RIMAP reproduces the large-scale trends and the main features of the solar wind plasma sampled by PSP, especially the crossing of the HCS. That is, the HCS-associated reversal of the background magnetic field, the null magnetic field intensity, and enhancements of the bulk speed and plasma density are all satisfactorily modelled by RIMAP. 
The major disagreements between PSP and RIMAP profiles in density and magnetic field happen where the agreement between input and output speed is worst, in particular in correspondence with the longitude intervals of [-5\textdegree;15\textdegree], [30\textdegree;55\textdegree], and [85\textdegree;110\textdegree]. 
With RIMAP, slower streams followed by high bulk speed gradients are not well reproduced, since many of them are removed during the back reconstruction in order to assure the physical consistency of the model \citep[see][]{Biondoetal2021}. Nevertheless, the proton density and the radial magnetic field show a good degree of accuracy in the reproduction of the main features encountered by PSP during the selected time interval, in particular during the HCS crossing at 65\textdegree, 115\textdegree, and 140\textdegree. It is finally worth noting that an equally good agreement between the model and observations cannot be obtained for the transverse components of the magnetic field, since its 3D treatment is still beyond the scope of our model, in which the reconstruction of $\Vec{B}$ is based on the classical, purely 2D description of \citet{weberdavis1967} and \citet{barker1982}. At the same time, PSP measurements show a transverse component of the magnetic field that is generally significantly smaller than the radial component and it fluctuates around $B_{\varphi}=0$. Our global Parker spiral reconstruction shows the same trend and thus it remains accurate even if it cannot reproduce the small-scale transients of the transverse magnetic field.


\begin{figure}[htbp]
\begin{center}
\includegraphics[width=\linewidth,trim={0.62cm 0.8cm 1.2cm 1.72cm},clip]{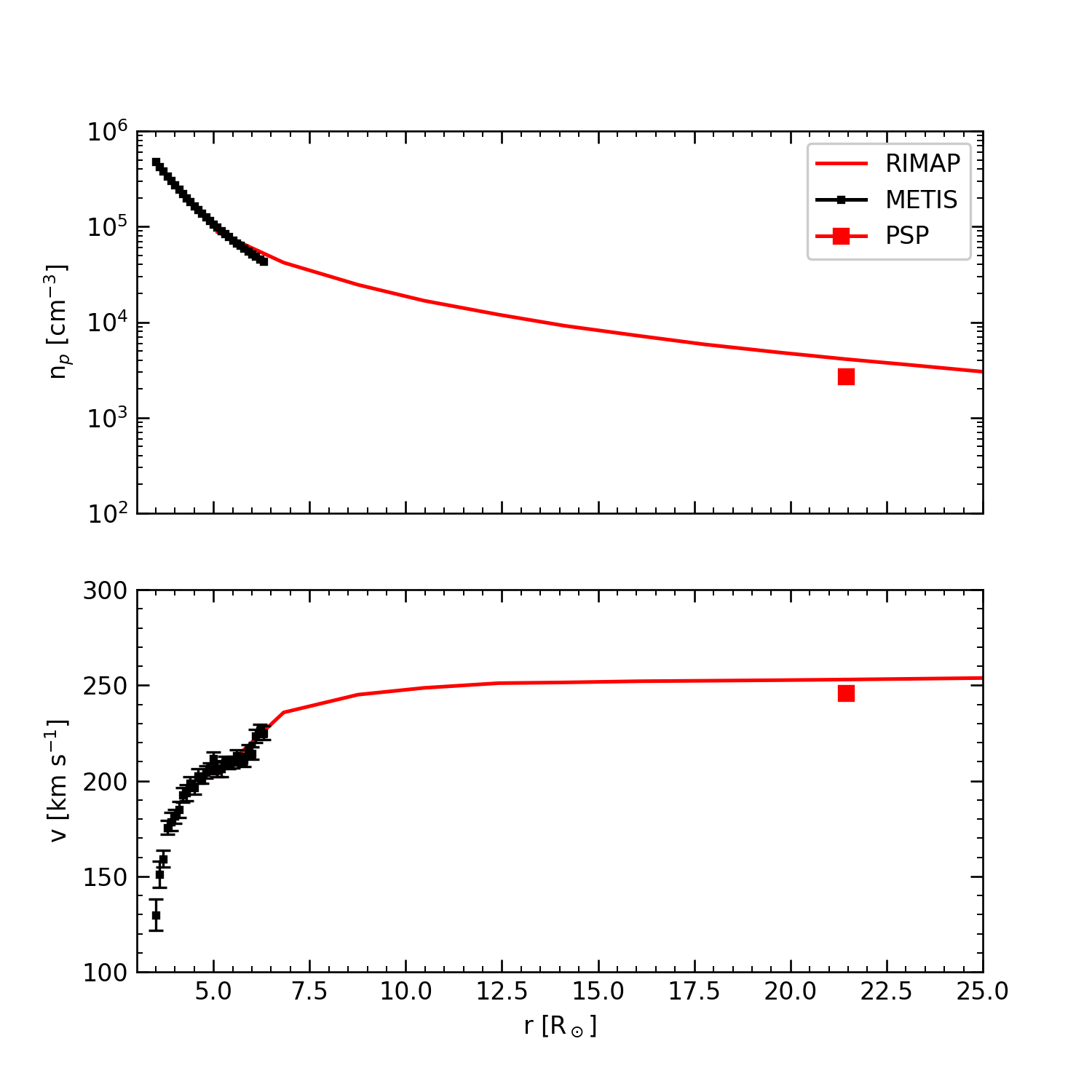}
\end{center}
\caption{Comparison between observed (squares) and RIMAP-modelled (red line) proton number density (top panel) and outflow velocity (bottom panel) along the same streamline, from $3$ to $25$ R$_{\odot}$. Black and red squares refer to Metis and PSP estimates, respectively.}
\label{fig:metis}
\end{figure}

In order to test the likelihood of the Parker spiral reconstructed by RIMAP from PSP data in the very inner heliosphere, model results extrapolated to the extended corona (i.e. for distances larger than $5$ R$_{\odot}$) were compared with outflow velocity and density estimates as inferred from Metis observations. On January 17, 2021, at 16:30 UT, the Metis coronagraph aboard Solar Orbiter (in quadrature with PSP at the east limb) indeed observed the solar corona in the field of view from $3.5$ to $6.3$ R$_{\odot}$. In particular, the Metis plane of the sky was at $99^\circ$ longitude. 
The electron density was derived by inverting the polarized brightness according to the technique developed by \citet{vandeHulst1950}. The coronal outflow velocity, on the other hand, was inferred from the intensity of hydrogen atoms, which is mainly, to a very first approximation, a function of density and velocity. Therefore, relying on the electron density derived as above \citep[and making plausible assumptions about the hydrogen temperature, helium abundance, and temperature anisotropy;][]{Dolei2016,Dolei2018}, a hydrogen intensity was synthesized as a function of only the expansion velocity of hydrogen atoms. From the comparison with the intensity observed by Metis, the speed of the outflowing plasma was then estimated. The interested reader is referred to Sect. 11 of \citet{antonucci2020} for more details on coronal diagnostic techniques and to \citet{telloni2021,telloni2022} for a more detailed description of how the speed and density of the coronal flow were derived.
Figure \ref{fig:metis} shows that the observed (black squares) and modelled (red curve) outflow velocity and proton number density along the same streamline \citep[a helium abundance of $2.5$ \% is assumed to relate electron to proton densities;][]{moses2020} are in striking agreement with each other, indicating that the RIMAP reconstruction is very reliable. On the other hand, the consistency of the model profiles with the data corroborates the calibration and reliability of Metis data processing.

After the steep radial trend of Metis observations close to the Sun, proton density begins to decline as $r^{-2}$, reaching approximately 3$\cdot10^3\text{ cm}^{-3}$ at 21.4 R$_\odot$, above the PSP measurement. In the bottom panel of Fig. \ref{fig:metis}, the plasma wind along the selected streamline starts with the sharp Parker-like acceleration close to the solar surface before attaining a constant speed of nearly 250 km/s, very close to the measured PSP one at 21.4 R$_\odot$.

\section{Conclusions}

Multi-spacecraft observations constitute a fundamental tool for fully understanding of solar phenomena and their dynamic evolution \citep{velli2020AA,hadid2021}. Among the possible configurations that can be assumed by two or more spacecraft, the quadratures provide a probe for remotely observing the coronal plasma that later impinges upon the other(s): this gives the opportunity to reconstruct the magnetic connectivity of the plasma observed in situ with its source near the Sun.

Through the Solar Orbiter-PSP quadrature of January 2021, \cite{telloni2021} followed for the first time the same plasma parcel of the solar wind from the sub-Alfvénic solar corona observed by the Metis instrument aboard Solar Orbiter to just above the Alfvén radius, where it was measured by PSP, and studied its magnetic connection relying on an empirical model.
A first theoretical modelling of the joint PSP-Metis/Solar Orbiter observation is presented in \citet{adhikari2022}, where the nearly incompressible MHD theory \citep{Zank_2017} is considered to successfully describe the slow solar wind flowing from the extended corona to the very inner heliosphere.
In this work, instead, we used the detailed hybrid analytical-numerical MHD model RIMAP to reconstruct the Parker spiral, binding it to PSP data collected from January 15-21, 2021. Then, we extracted the streamline passing through the intersection of the PSP trajectory with the longitude of the Metis plane of the sky on January 17. In this way, the plasma properties detected in the lower solar corona by PSP are magnetically connected to those observed by Metis immediately above the solar surface via a detailed and data-driven MHD simulation.

This work required improving the model over its previous iterations \citep{Biondoetal2021}. RIMAP can now ingest input data from spacecraft with pronounced-eccentricity orbits (i.e. non-constant heliocentric distances and irregular $d\varphi$). In addition, improvements have been made in the construction of the internal boundary, which now takes the acceleration experienced by the solar wind during the expansion of the streamlines into account, conveniently reducing the speed in the back-mapping phase.

The satisfactory agreement between Metis remote-sensing observations and PSP in situ measurements confirms not only the validity of RIMAP Parker spiral reconstructions, but also the possibilities offered by multi-spacecraft observations when investigating heliospheric physics through the refinement of MHD simulations at unprecedented levels of detail. This accuracy will likely improve as PSP and Solar Orbiter approach distances closer to the Sun, especially when the former enters the solar corona.

Further applications of RIMAP will include the PSP-Solar Orbiter quadrature of June 2022, during which PSP was just 0.4 R$_\odot$ out of the Metis field of view and, likely, in the sub-Alfvénic regime. Our model could then be used to reconstruct remote-sensing and in situ measurements of the solar corona. More generally, accurate reconstructions of the internal Parker spiral bound with PSP data could provide an adequate test bench for the study and modelling of the propagation of transient phenomena and MHD fluctuations, such as turbulence and switchbacks.\\

Solar Orbiter is a space mission with an international collaboration between ESA and NASA, operated by ESA. Metis was built and operated with funding from the Italian Space Agency, under contracts to the National Institute of Astrophysics (INAF) and industrial partners. Metis was built with hardware contributions from Germany (Bundesministerium für Wirtschaft und Energie through DLR), the Czech Republic (PRODEX) ,and ESA. PSP data were downloaded from the NASA’s Space Physics Data Facility (https:// spdf.gsfc.nasa.gov).
\bibliographystyle{aa} 
\bibliography{rimap3} 
\end{document}